# Synthesis of infinite-layer nickelates and influence of the capping-layer on magnetotransport

*Guillaume Krieger[1], Aravind Raji[2], Laurent Schlur[1], Gilles Versini[1], Corinne Bouillet[1], Marc Lenertz[1], Jérôme Robert[1], Alexandre Gloter[2], Nathalie Viart[1] and Daniele Preziosi[1,\*]*

[1] Université de Strasbourg, CNRS, IPCMS UMR 7504, F-67034 Strasbourg, France
[2] Laboratoire de Physique des Solides, CNRS, Université Paris-Saclay, 91405 Orsay, France

**Abstract**

The recent discovery of a zero-resistance state in nickel-based compounds has generated a re-excitement about the long-standing problem in condensed matter of high-critical-temperature superconductivity, in light of the analogies between infinite-layer nickelates and cuprates. However, despite some formal valence and crystal symmetry analogies, the electronic properties of infinite-layer nickelates are remarkably original accounting, among other properties, of a unique Nd5d-Ni3d hybridization. This designates infinite-layer nickelates as a new class of oxide superconductors which should be considered on their own. Here we report about $Nd_{1-x}Sr_xNiO_2$ (x = 0, 0.05 and 0.2) thin films synthesized with and without a $SrTiO_3$ capping-layer, showing very smooth and step-terraced surface morphologies. Angle-dependent anisotropic magnetoresistance measurements performed with a magnetic field rotating in-plane or out-of-plane with respect to the sample surface, rendered important information about the magnetic properties of undoped $SrTiO_3$-capped and uncapped samples. The results point at a key role of the capping-layer in controlling the magnitude and the anisotropy of the anisotropic magnetoresistance properties. We discuss this control in terms of a combined effect between the Nd-Ni hybridization and an intra-atomic exchange coupling between the Nd-4f and Nd-5d states, the latter essentially contributing to the (magneto)transport. Further studies foresee the influence of the capping layer on infinite-layer nickelates with no magnetic rare-earth.

**Introduction**

The observation of superconductivity in infinite-layer $Nd_{0.8}Sr_{0.2}NiO_2$ thin films[1] has finally put an end to more than 20-years of sustained experimental efforts devoted at mimicking the hallmarks of cuprates superconductivity. This recent discovery has re-vitalized the quest of new high-Tc materials with unusual $Ni^{1+}$ and square-planar coordination, as initially suggested by a seminal theoretical work[2]. Soon after the first report, La- and Pr-based infinite-layer nickelates[3,4], and reduced square-planar Ruddlesden-Popper $Nd_6Ni_5O_8$[5] thin films have also shown a zero-resistance state, assessing the importance of nickelates as a material platform to engineer novel superconductors.

From the preparation standpoint the two-step synthesis process[6] is very challenging, as explained later and, so far, only a handful of works[7,8,9] showed that it could be independently reproduced. The recipe provides (I) the optimization of the Sr-doped perovskite phase, with the subsequent, (II) $CaH_2$-assisted topotactic reduction. The tetragonal infinite-layer phase (*a* = 0.392 nm, *c* = 0.328 nm as reported for the bulk undoped parent compound[10]), shows a very small lattice mismatch (*ca.* -0.3 %) with

SrTiO$_3$ (STO) ($a$ = 0.391 nm). On the contrary, the perovskite phase ($a_{pc}$ = 0.381 nm for NdNiO$_3$), experiences a very large lattice mismatch (*ca.* +2.5 %) with STO, which poses severe problems for its stabilization due to an already proven increase of the oxygen vacancies concentration for tensile strains[11]. This, combined to possible Nd/Ni off-stoichiometry issues[12,13], could trigger an undesirable presence of Ruddlesden-Popper-like phases, hindering the stabilization of the infinite-layer phase during the topotactic reduction in the entire volume of the sample. Additionally, upon Sr-doping the precursor Nd$_{1-x}$Sr$_x$NiO$_3$ phase requires a further fine balance of the growth parameters to avoid Sr-segregation[14], and to stabilize the large Ni valence state (Ni$^{3+x}$) forced beyond the already unstable 3+ value. Basically, a high density of extended defects mixed to secondary phases may undermine the appearing of superconductivity[6]. Indeed, a superconducting state was also reported for Sr-doped LaNiO$_2$ thin films[4,15], although pioneering attempts failed[1], pointing *de facto* at the major role played by the sample quality. Significant advances in thin films crystallinity, as obtained for growth onto (LaAlO$_3$)$_{0.3}$(Sr$_2$TaAlO$_6$)$_{0.7}$ (LSAT) ($a$ = 0.387 nm) single crystals[16], largely increased the superconducting dome with onset critical temperatures above 20 K.

As a matter of fact, it is worth noting that undoped NdNiO$_2$ thin films could not be stabilized onto LSAT. The fact that, up to date, this could not be reproduced by the same groups that, on the other side, managed to get superconductivity for STO-grown samples, tells that the synthesis process of infinite-layer nickelates is, indeed, very delicate and not yet fully understood. On a purely experimental basis, it seems necessary to adapt the topotactic reduction parameters sample by sample, since, for different Sr-contents and/or rare-earth ions, they can be characterized by a different concentration of extended defects. One may argue that a limited number of "extended defects" is necessary to perhaps provide a preferential path for the oxygen ions to migrate.

It was originally suggested that an STO capping-layer could allow for a better stabilization of the infinite-layer phase during the topotactic reduction[1]. Beyond the assumption that the capping-layer grants a protection from the direct contact with the reducing agent, thus minimizing decomposition problems[6], its role is still unclear. Indeed, superconductivity has also been reported for uncapped films, suggesting that it is not due to the presence of the capping-layer[7], and some properties strongly differ for capped and uncapped samples, in particular the low-energy excitation features of resonant inelastic X-ray scattering (RIXS) measurements, as some of the authors already reported in a previous work[17]. Those existing challenges regarding the growth of Sr-doped perovskite nickelate thin films, combined with the pivotal handling of the topotactic reduction (fine balance among temperature and time of reaction), make it hard to establish a reproducible route towards superconducting infinite-layer nickelates.

In this work Nd$_{1-x}$Sr$_x$NiO$_2$ (x = 0, 0.05 and 0.2) thin films were synthesized with and without a SrTiO$_3$ capping-layer with similar structural and transport properties as already reported in literature and exhibiting a never reported so far step-terraced morphology. The superconducting Nd$_{0.8}$Sr$_{0.2}$NiO$_2$ samples were, nevertheless, characterized by a residual resistance value at 2 K. Angle-dependent anisotropic magnetoresistance (AMR) measurements for undoped capped samples rendered a modified two-fold symmetry and stronger intensities than the ones observed for the uncapped samples. The magnetotransport results allow to catch a glimpse on the relation between the intrinsic magnetism of NdNiO$_2$ thin films mainly dictated by the presence of the Nd 4f states, and perhaps by the Nd-Ni hybridization modified by the presence/absence of a STO capping-layer[17]. The paper is organized as follows: after the experimental section, we introduce the study concerning the growth of perovskite Nd$_{1-x}$Sr$_x$NiO$_3$ thin films encompassing structural and transport properties. Then, in the second part we present the topotactic reduction study followed by a third section dedicated to the angle-dependent AMR results.

## Experimental Section

Thin films have been grown by pulsed laser deposition (PLD), assisted by a reflection high-energy electron diffraction (RHEED) technique to monitor *in situ* the growth of the Sr-doped perovskite nickelates thin films. Undoped NdNiO$_3$ thin films were obtained by ablating a single-phase ceramic target, while the Sr-doped films were obtained by ablating ceramic targets composed by a stoichiometric mixture of NiO and Sr-doped Nd$_2$NiO$_4$ phases (Toshima Manufacturing Co.). The latter represents the n = 1 member of the Ruddlesden-Popper (RP) R$_{n+1}$Ni$_n$O$_{3n+1}$ nickelate family, which is the layered analogue of the perovskite one. 5x5 mm² STO (001) substrates (CODEX-International) were used as growth templates, which underwent the Kawasaki procedure[18] to present a well-defined step-terraced surface. 5x10$^{-8}$ mbar was the usual base pressure of the PLD chamber prior deposition, and all the growth experiments were performed with the relatively large 0.3 mbar value for the oxygen partial pressure, hereafter indicated as P(O$_2$). A KrF excimer laser ($\lambda$=248 nm) was operated at a repetition rate of 2 Hz, and we used two different ablation conditions depending on the type of the targets, *i.e.* single and/or mixed phases. In particular, highly crystalline undoped NdNiO$_3$ [hereafter indicated as NSNO(0)], thin films could be obtained by scanning the laser over a few square millimetres surface area of the ceramic target, while for the Nd$_{0.95}$Sr$_{0.05}$NiO$_3$ [NSNO(5)] and Nd$_{0.8}$Sr$_{0.2}$NiO$_3$ [NSNO(20)] thin films, the laser was maintained on the same spot during the entire ablation process (no-toggling). At the end of the deposition which was performed at the temperature of 675 °C the samples were cooled down to room temperature at a rate of 5 °C/min under the same growth oxygen partial pressure. Each different sample was obtained ablating a new target zone with relatively high fluence values to mitigate some possible off-stoichiometry issues to which nickelates are highly sensitive as already pointed out by some of the authors[13]. For the capped samples an STO single crystal was ablated to get a 3-unit-cells (u.c.) thick capping layer. We used a Rigaku Smartlab diffractometer equipped with a rotating anode and a monochromated copper radiation ($\lambda$ = 0.154056 nm) for the X-ray diffraction (XRD), X-ray reflectivity (XRR) and reciprocal space mapping (RSM) measurements. The thickness of the samples prior and after topotactic reduction were obtained by fitting the XRR curves via the GlobalFit software. The surface morphologies were acquired by using a Park XE7 (Park System) Atomic Force Microscope (AFM) in true non-contact-mode, and their root-mean-square roughness values were quantified by the XEI software from Park System. Cross-sectional Transition Electron Microscopy (TEM) lamellae were prepared using a focused ion beam (FIB) technique (D. Troadec at IEMN facility Lille, France; and at C2N, university Paris-Saclay, France). They were observed in a NION Ultra-STEM200 $C_3/C_5$-corrected scanning transmission electron microscope (STEM), using high-angle annular dark-field imaging (HAADF) and Electron Energy Loss Spectroscopy (EELS). The transport properties of the films were studied with a Dynacool system (Quantum Design) equipped with a sample-rotator. The samples were cut in a 2.5x2.5 mm² geometry, and measured in a van der Pauw method by applying current amplitudes of 10 µA. The rotation took place in the 0°-360° range in two distinct geometries, hereafter referred to in-plane (IP) and out-of-plane (OOP). For the in-plane geometry the rotation of the magnetic field (H) takes place within the $NiO_2$ plane from the H∥I parallel configuration ($\varphi = 0°$), towards the perpendicular one H⊥I ($\varphi = 90°$). On the other hand, for the out-of-plane geometry H rotates in a plane perpendicular to I (null Lorentz magnetoresistance configuration), from the in-plane ($\theta = 0°$) towards the out-of-plane ($\theta = 90°$) direction with respect to the sample surface (see Fig. 12a).

## Growth of Sr-doped perovskite nickelate thin films

Bulk $NdNiO_3$ exhibits an orthorhombic crystal structure with a lattice parameter of *ca.* 0.381 nm in pseudo-cubic notation. Thus, the large tensile strain experienced onto STO single crystals when grown as thin film produces a contraction of the out-of-plane lattice parameter (***c***-axis), which nominal value is expected to be equal to 0.375 nm ($2\theta \sim 48.5°$), accounting for a Poisson ratio[19] of 0.3. Departure from this ***c***-axis value can be used as a proxy to highlight possible oxygen deficiencies, which not only provoke a lattice expansion, as demonstrated in similar PLD-grown $NdNiO_3$ thin films[11], but also modify the unstable $Ni^{3+}$ valence landscape, by creating defect-zone where the unitary Nd/Ni stoichiometry level is not overall conserved. This direct link between ***c***-axis elongation and off-stoichiometry issues may become more complicated to make in the case of Sr-doping for which an elongation of the ***c***-axis is, indeed, expected as a result of the chemical substitution with a cation of a larger ionic radius. As a result, a study of the influence of the oxygen partial pressure combined to the respective transport behaviour is a necessary step to ascertain that a 'pure phase' of Sr-doped perovskite nickelate thin films can be, indeed, stabilized. Therefore, the out-of-plane cell parameter deduced from the $2\theta$ peaks, combined to the transport properties give important information about Nd/Ni off-stoichiometries and/or oxygen vacancies[12]. Garcia *et al.* have already showed that hole-doping bulk rare-earth nickelates by chemical substitution, results in a gradual closing up of the charge-transfer gap. As a result, the usually observed metal-to-insulator transition (MIT) for undoped samples is heavily modified, and at already 5% at. Sr doping the samples exhibit an overall metallic behaviour[20].

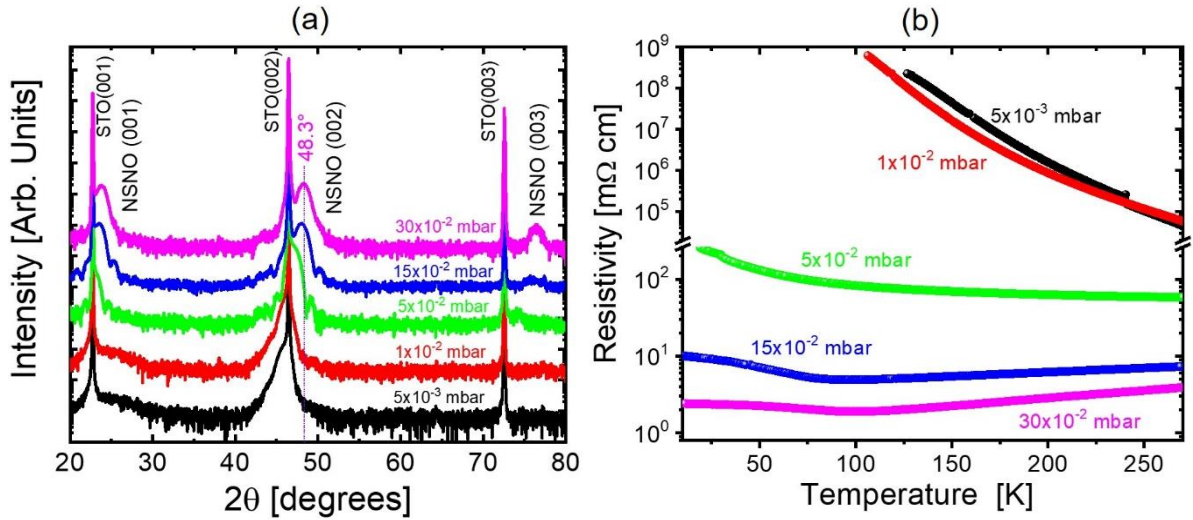

*Figure 1 (a) XRD and (b) Resistivity measurements of NSNO(5) thin film grown with different oxygen partial pressure values.*

Figure 1a shows XRD patterns of uncapped NSNO(5) thin films grown at different $P(O_2)$ values, with a laser fluence of 3.8 J/cm². The number of laser pulses were adjusted to account for the $P(O_2)$-modified growth conditions to guarantee the same thickness value of *ca*. 10 nm for each sample. For the lowest value of $5\times10^{-3}$ mbar (black curve), the only visible shoulder on the left of the STO 002 reflection, points at a not properly crystallized sample. Due to the very low oxidizing conditions, this shoulder can be ascribed to $Ni^{2+}$-based phases, such as NiO and/or Sr-doped $Nd_2NiO_{4-\delta}$[21]. By increasing $P(O_2)$ for consecutive growth experiments this peak gradually shifts towards higher $2\theta$ values until it reaches the 48.3° position for 0.3 mbar (magenta curve). At this $P(O_2)$ value we can also easily observe a relatively intense NSNO(5) 003 reflection which, according to our experience, highlights that the perovskite nickelate phase is stabilized at a good degree level. Transport measurements are then necessary to demonstrate that a 'pure' metallic state could be stabilized as indeed expected for a 5% at. Sr doping. Figure 1b shows the temperature dependence of the resistivity for the entire uncapped NSNO(5) sample series. As expected, we observed a very different behaviour for the various $P(O_2)$ values. Thin films grown at a very low oxygen partial pressure exhibit a fully insulating behaviour typical of $Ni^{2+}$-based compounds, such as RP phases[22] or NiO[23]. As $P(O_2)$ reaches values for which a degree of oxidation close to $Ni^{3+}$ is guaranteed, we observe a metallic behaviour at room temperature and an upturn below 100 K, less pronounced when keeping increasing the oxygen partial pressure. Usually the upturn in resistivity is explained as the result of charge localization triggered in this particular case by the presence of extended defects within the thin films. At 0.3 mbar the sample is characterized by an overall lower resistivity, and the still present small upturn points at a non-properly optimized growth. As already showed in literature, possible off-stoichiometry issues in perovskite nickelate thin films are largely due to the impossibility to correctly transfer the Ni from the target towards the substrate, and therefore, the target ablation history combined to the laser fluence become important growth parameters[13]. On this basis, we resorted to a laser ablation onto a target which is kept fixed (no-toggling condition), and increased the laser fluence to 4.8 J/cm², an approach that gave the proper leverage over the Nd/Ni stoichiometry degree in the laser-induced plume. Figure 2a shows the comparison between two XRD patterns acquired for uncapped NSNO(5) samples prepared with these two different approaches. With the no-toggling condition, and at an higher fluence than the one used for NSNO(0), it has been possible to improve the quality of the Sr-doped thin films as indicated by a larger intensity of the peaks which are now more separated from the STO ones pointing at a smaller *c*-axis parameter. Parallel to this, a further decrease of the overall resistivity of the samples has been observed with no upturn as shown in Figure 2b, hence, signalling an improved crystallinity. The same results hold also for NSNO(20) samples as shown in Figures 3a,b.

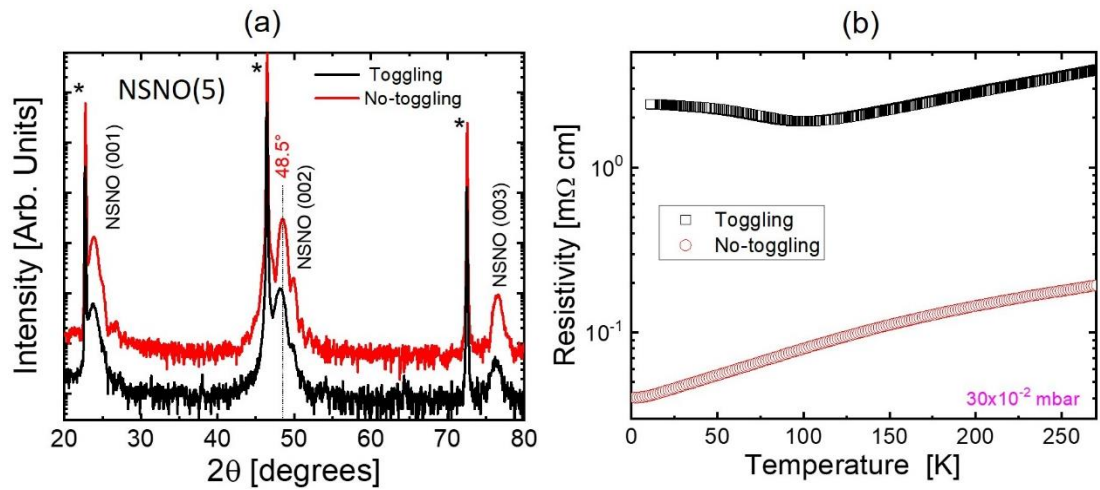

*Figure 2 (a) XRD and (b) Resistivity measurements of NSNO(5) samples prepared with toggling and no-toggling conditions.*

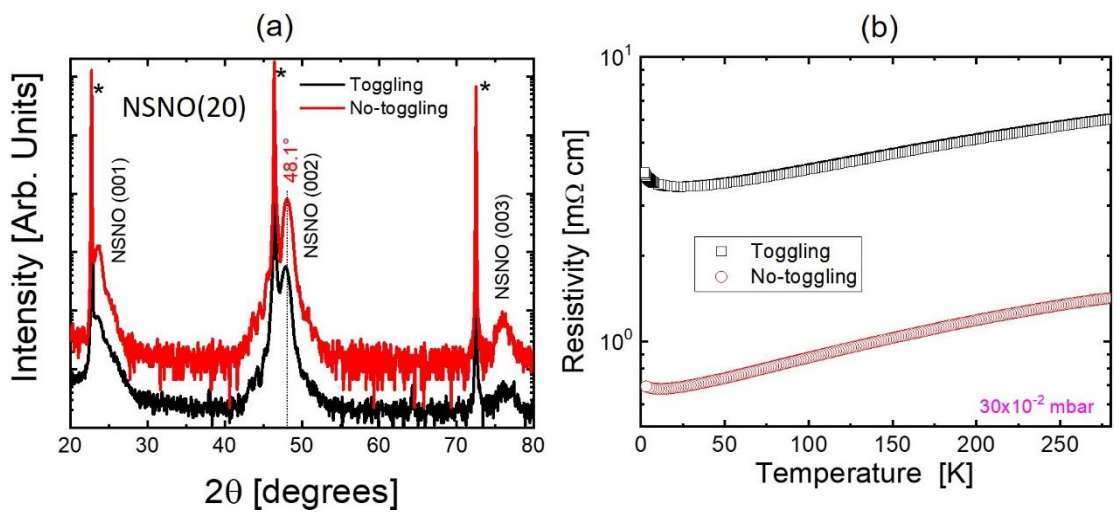

*Figure 3 (a) XRD and (b) Resistivity measurements of NSNO(20) samples prepared with toggling and no-toggling conditions.*

Although the overall resistance for the NSNO(20) samples decreased when resorting to a no-toggling ablation condition, one must note that, its value is nevertheless higher than the one measured for the optimized NSNO(5) samples. This point at a still relatively high number of defects for the NSNO(20) samples.

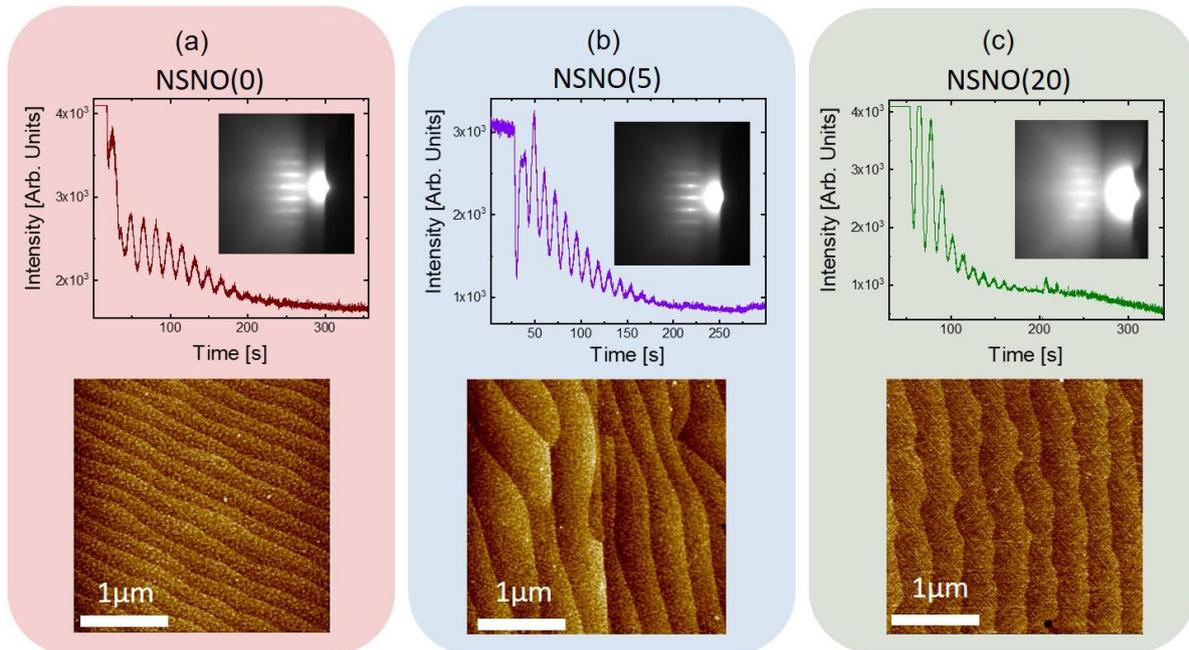

*Figure 4 (Top part) Temporal evolution of the RHEED intensity during the growth with some initial oscillations. The insets show the final RHEED pattern showing an overall 2D character. (Bottom part) AFM images showing a good step-terraced image mimicking the STO surface profile. The rms-roughness values varying within the 0.2-0.3 nm range, for (a) NSNO(0), (b) NSNO(5) and (c) NSNO(20) samples synthesized with optimized growth parameters*

Figure 4 summarizes the surface characterizations of the Sr-doped perovskite thin films obtained by monitoring the intensity of the RHEED oscillations and patterns during growth (upper part of Fig. 4), and afterwards by AFM measurements (lower part of Fig. 4). Very clear oscillations of the RHEED intensity could be observed for all the samples at the very beginning (estimated growth rate ranging between 25-30 pulses/u.c. depending upon the Sr-content), and then get lost as the deposition proceeds. The impossibility to observe the RHEED oscillations until the end of the deposition experiments (overall 700-1000 laser pulses), could be ascribed to the optimized P($O_2$) and substrate-temperature values used for the growth (see Table 1), and necessary to properly stabilize the $Ni^{3+x}$ valence state. In particular, this combination of growth parameters may set a non-optimal kinetic energy of the ad-atoms crucial to properly diffuse and fully cover the substrate's terraces, which then experience an incremental increase of their roughness level. This is most likely the cause of the appearing of detrimental secondary and/or Sr-segregated phases which manifest as a 2D-3D transition of the RHEED pattern. Indeed, according to our experience thicker Sr-doped thin films (>15 nm) exhibit a fully 3D RHEED pattern, and the usual metallic behaviour is replaced by an insulating state which poses severe limitations to the maximum allowed thickness for this system. On this basis, we have calibrated our deposition time by monitoring the appearing of those 3D-like features in the RHEED patterns, which keeps an optimal 2D degree level for film thicknesses equal to *ca.* 10 nm, as it is possible to infer by a close inspection of the insets of Figure

4. Finally, our 10 nm thick Sr-doped perovskite samples show atomically flat AFM images with a step-and-terrace morphology typical of a coherent layer-by-layer growth.

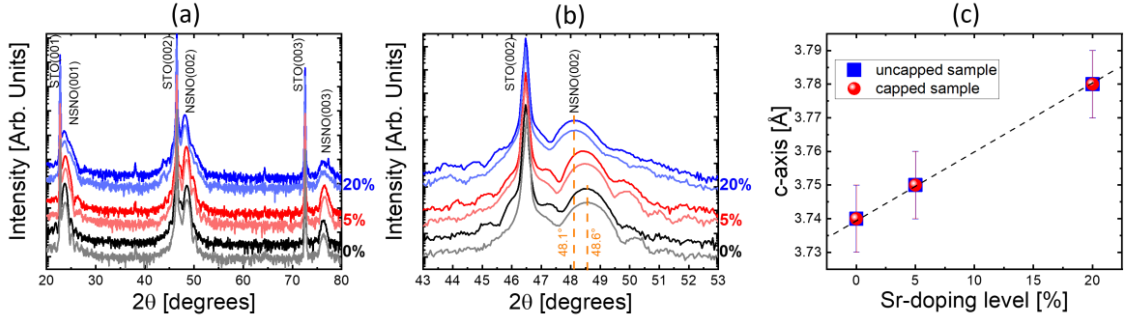

*Figure 5 (a-b) XRD for the three different doping levels of perovskite samples, with and without a capping layer. Bright colours are for the capped samples, and light colours for the uncapped samples. (b) Zoom on the (002) peaks. (c) Linear dependence of the c-axis parameters as a function of the Sr doping and capping.*

As a summary of the structural properties we present in Figure 5a a comparison among the XRD patterns acquired for the perovskite NSNO(x) sample series (x = 0, 0.05 and 0.2) with (bright-colour) and without (light-colour) STO capping layer, and grown by using the optimized parameters listed in Table 1. All the (00L) XRD peaks with L = 1,2,3 are present, and showing a large intensity which speaks in favour of a high crystallinity degree of the samples. Moreover, the presence of the (003) peak for all samples can be used as a proxy of a good stabilization of the perovskite phase as already pointed out in the discussion of Figure 1. The zoomed area taken around the NSNO(x) (002) peak position, and presented in Figure 5b, beyond showing an increase of the out-of-plane lattice parameter as a function of the Sr-doping, demonstrates that the relatively thin capping layer of STO (3 u.c.), does not affect the XRD patterns from a macroscopic point of view. This information is displayed in Figure 5c where we can easily infer that the increase of the Sr-content brings to a linear increase of the *c*-axis parameter with the same values for capped and uncapped samples, within the experimental error.

| Target | Thin film | Fluence | Temperature | Oxygen | Ablated area |
|---|---|---|---|---|---|
| $NdNiO_3$ | $NdNiO_3$ | 3.8 J/cm² | | | 5x5 mm² |
| $Nd_{1.9}Sr_{0.1}NiO_4$+NiO | $Nd_{0.95}Sr_{0.05}NiO_3$ | 4.8 J/cm² | 675 °C | 0.3 mbar | No toggling |
| $Nd_{1.6}Sr_{0.4}NiO_4$+NiO | $Nd_{0.8}Sr_{0.2}NiO_3$ | | | | |
| $SrTiO_3$ single crystal | $SrTiO_3$ | 2.4 J/cm² | | | 3x3 mm² |

*Table 1: Optimized growth parameters for the perovskite nickelate samples series.*

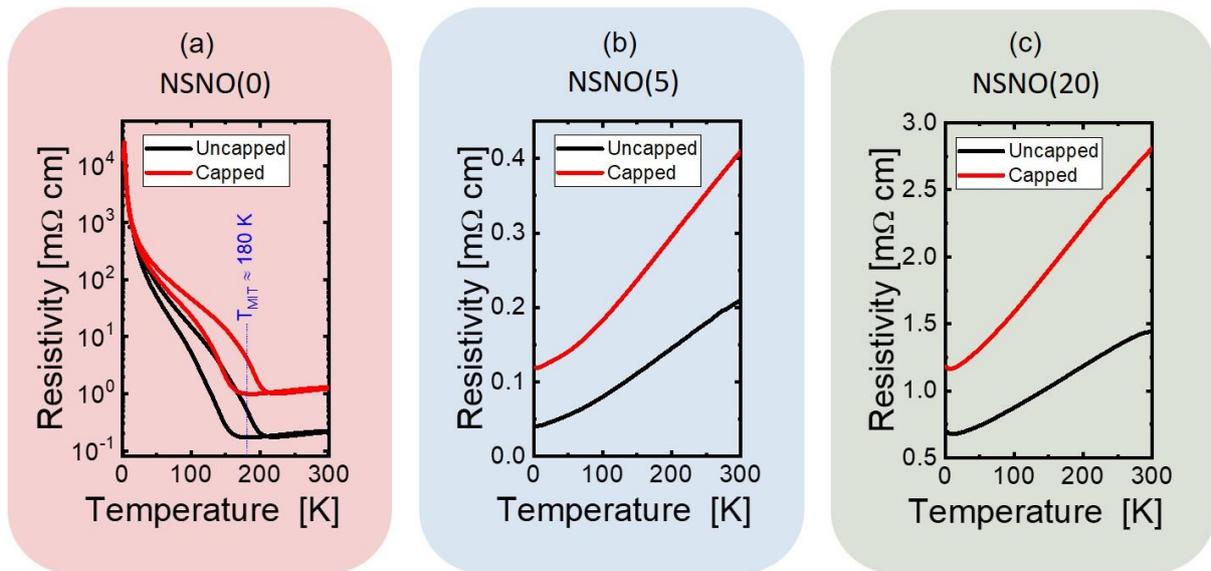

*Figure 6 Temperature dependence of the resistivity for the three different doping levels of the perovskite samples, with and without STO capping-layer.*

To conclude this first part dedicated to the growth optimization of the perovskite Sr-doped nickelate thin films we show in Figure 6 a comparison of the transport properties for both capped and uncapped sample series. The only 3 u.c. thick STO capping layer does not alter the temperature onset of the Metal-to-insulator Transition (MIT) for undoped samples which show same values as bulk[24]. Upon Sr-doping the MIT is fully destroyed as already discussed before, and we observe a relative higher resistivity value for capped samples. It is worth noticing that $R$NiO$_3$ ($R$ being a rare-earth), have been studied since decades for their sharp MIT properties[24]. The rich phase diagram of bulk $R$NiO$_3$ displays an intimate connection between the Ni-O-Ni bond angles/lengths (controlled by the choice of $R$), and the temperature at which the MIT is encountered ($T_{MIT}$). The modulation of the MIT through epitaxial strain and/or electron/hole-doping in thin films has been a productive field of research[25–28]. The oxygen vacancies may also play a decisive role in the measured $T_{MIT}$ values by locally altering the Ni-O-Ni network path. This possibility is largely enhanced in the presence of an oxide capping layer, as in this particular case. Indeed, if not properly optimized the growth of the STO capping layer can alter the interfacial Ni-O-Ni bond angles/lengths, and together with a modified Ni valence state cause important variation of the transport properties.

**Synthesis of infinite-layer nickelate thin films**

First topotactic reduction processes for LaNiO$_3$ in both bulk[29] and thin film[30,31] forms were performed by using a delicate H$_2$ gas-flow set-up. Resorting to binary metal hydrides, such as NaH[10,32] and CaH$_2$[1,33–35], had the big advantage of largely simplifying the experimental approach. Kobayashi *et al.*[36] found that the reduction is more efficient when the powder is in contact with the precursor, although the underneath mechanisms controlling the reaction are not yet fully understood. We have performed the topotactic reduction by using an evacuated silica tube sealed via a membrane-valve as shown in the inset of Figure 7b, and in which the 0.5 g of CaH$_2$ (Sigma Aldrich) powder was put in direct contact with our 10 nm thick Nd$_{1-x}$Sr$_x$NiO$_3$ films.

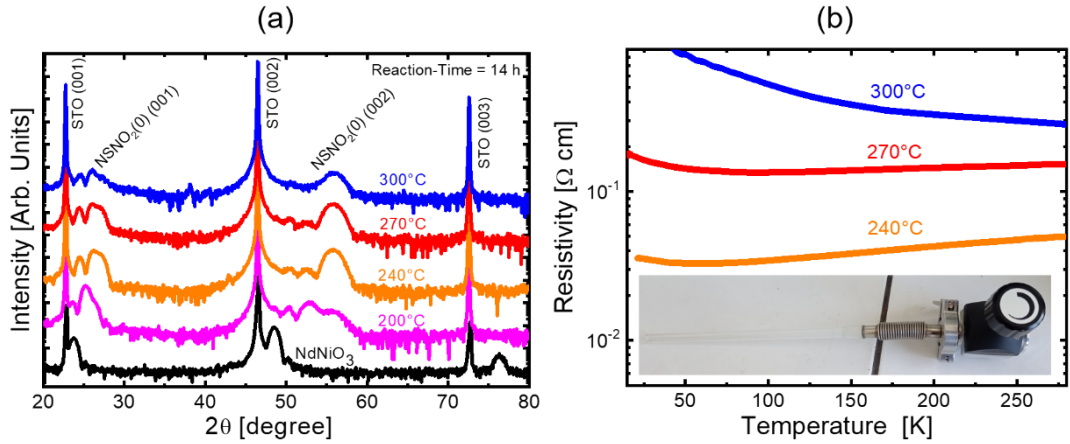

*Figure 7 (a) XRD of NSNO(0) thin films after different topotactic reduction processes with (b) related transport properties. The inset shows the set-up used for all the oxygen de-intercalation experiments.*

Figures 7a,b show the temperature study of a series of topotactic reduction processes (14 hours long) undertaken for uncapped NSNO(0) samples. At 200 °C the infinite-layer phase is not properly stabilized since the shift of the XRD peaks is only partial. At higher temperatures of 240 °C and 270 °C the position and intensity of the XRD peaks signal a proper crystallinity of the infinite-layer phase with the expected out-of-plane lattice parameter[6]. Further increasing the temperature has no relevant effect on the position of the peaks, but the diminished intensity marks a worsening of the infinite-layer crystallinity (300 °C data). This is also confirmed from the overall insulating properties observed for this temperature (Fig. 7b). On the other side, at 240 °C and 270 °C we observed the expected transport properties for undoped samples, *i.e.* metallic followed by a small upturn around 70 K. In Fig. 7b we can easily infer that the 240 °C reduction temperature leads to an overall lower resistance for the infinite-layer phase, and therefore, we kept this value for all our NSNO(0) reduction experiments. Interestingly, the found set of parameters (14 hours and 240°C) gave similar results for NSNO(5) thin films. For NSNO(20) thin films we observed a superconducting transition only for a modified set of parameters, *i.e.* 3 hours and 270°C. Moving to the capped NSNO(0) and NSNO(5) samples, a reduction time of 30 hours was necessary to observe the expected shift of the XRD peaks combined to the optimal transport properties. Again, to observe a signature of the superconducting transition for our capped NSNO(20) samples, a higher reduction temperature value than the one used for the uncapped samples was necessary. Table 2 summarizes the topotactic reduction parameters for both NSNO(x) capped and uncapped series of this work.

| Precursor phase | STO Capping-layer | Temperature [°C] | Time [h] |
|---|---|---|---|
| NdNiO$_3$ and Nd$_{0.95}$Sr$_{0.05}$NiO$_3$ | No | 240 | 14 |
| | Yes | 240 | 30 |
| Nd$_{0.8}$Sr$_{0.2}$NiO$_3$ | No | 270 | 3 |
| | Yes | 300 | 3 |

*Table 2 Optimized temperatures and durations for the topotactic reduction processes as a function of the different Sr-doping levels, and the presence/absence of the STO capping-layer.*

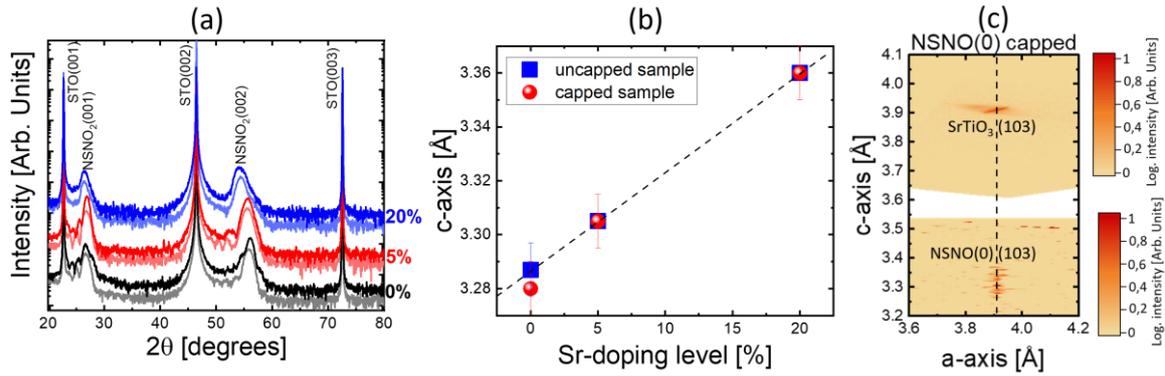

*Figure 8 (a) XRD for the three different doping levels prepared with and without capping layer. Bright colours are for the capped samples, and light colours for the uncapped ones. (b) linear dependence of the c-axis parameter as a function of the Sr doping and capping. (c) Reciprocal space maps of the STO (103) and NSNO (103)$_{pc}$ reflections for a capped NSNO(0) thin film, showing that the infinite-layer is fully strained to the substrate.*

As a summary of the structural properties of our infinite-layer thin films, we present in Figure 8a the XRD patterns acquired for NSNO(x) samples prepared with (bright colours) and without (light colours) STO capping-layer. As it is possible to infer from Figure 8b the *c*-axis parameter increases linearly with the Sr-doping level, and are not modified (within the experimental error), by the presence of the capping-layer. The NSNO(x) samples are fully strained to the substrate after the reduction as it is possible to infer from the RSM shown in Figure 8c.

So far, it has been reported that topotactic reduction processes damage the surface of the precursor phase, and a post-reduction vacuum annealing was mandatory to observe step-terraced morphologies[37]. The topotactic reductions presented here, performed at relatively low temperatures, granted morphologies of the infinite-layer phase preserving the smooth surface of the precursors. Figure 9 summarizes the AFM study of the reduced NSNO(x) capped series characterized by a rms-roughness value ranging in the 0.3-0.4 nm.

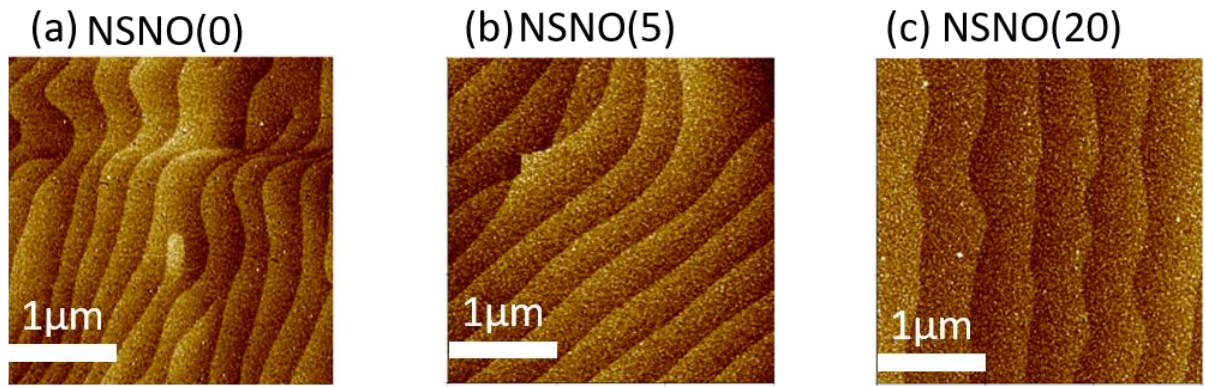

*Figure 9 3x3 µm² AFM images of reduced (a) NSNO(0), (b) NSNO(5), and (c) NSNO(20) samples, characterized by 0.3-0.4 nm rsm-roughness values.*

Figure 10 shows the temperature dependences of the resistivity curves acquired after reduction for our NSNO(x) capped and uncapped sample series. The transport properties of NSNO(0) and NSNO(5) are representative of the ones reported in literature[1,7], for both capped and uncapped cases. NSNO(20) samples show the expected superconducting transition around 10 K, but without reaching the zero-resistance state down to 2 K. This is most likely ascribable to a presence of relatively large density of extended defects in the precursor phase, combined to a still poor optimization of the reduction parameters for the 20% Sr-doped samples. The inset of Figure 10c shows the temperature dependence of the Hall coefficients exhibiting the same behaviour as the one reported in previous works[1,7]. To notice the sign change of the majority charge carriers around 50 K in the case of the superconducting samples (from electrons to holes).

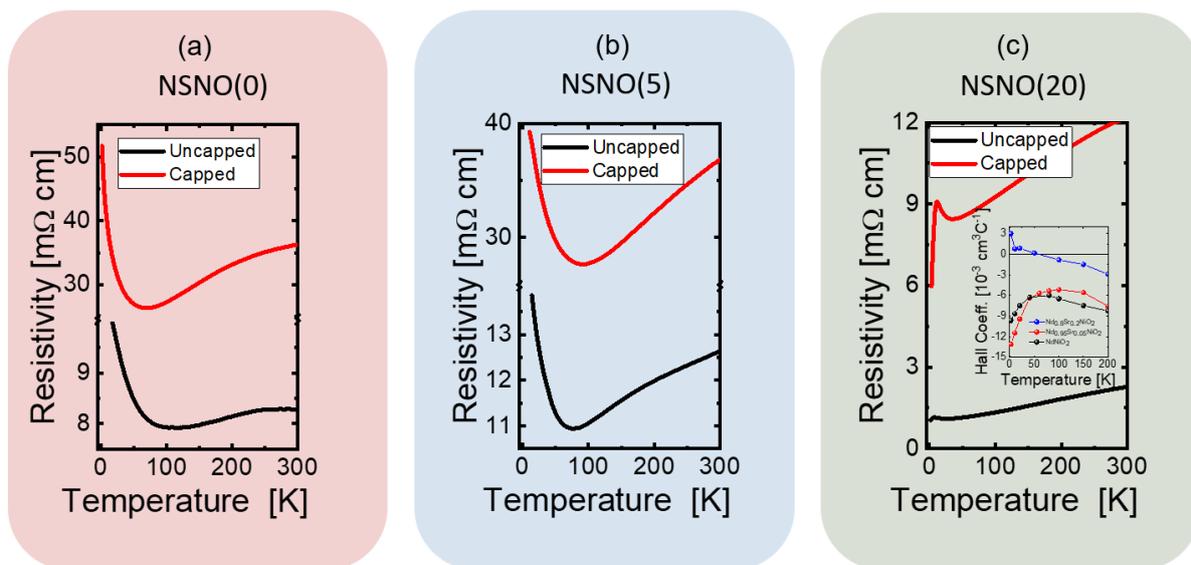

Figure 10 Resistivity curves of the infinite-layer samples as a function of temperature, for the three different doping levels, with and without an STO capping-layer. Inset shows the Hall effect measurements.

To conclude this second part, Figure 11 shows the STEM and EELS measurements of a reduced uncapped NSNO(0) sample which has been exposed to air. The atomically resolved HAADF-STEM image of the Figure 11a clearly demonstrates the high crystallinity of the NSNO(0) with a coherent growth onto the STO substrate. The Fast Fourier Transform (FFT) of different regions of the film are shown in Figure 11e. The *c*-axis parameter of the centre (R2) and substrate interface (R3) regions is of 0.33 nm, in agreement with the infinite-layer lattice parameter, and is getting larger at the surface of the thin film (R1) with a value close to 0.36 nm. The elemental profiles have been obtained from the Ni-L and Nd-M EELS edges (Figure 11b). The elemental quantification from calculated EELS cross sections have low accuracy when comparing edges of different shells (Figure 11c). We had first estimated experimentally the Ni-L / Nd-M EELS cross-section ratio on a reference $NdNiO_3$ thin film and applied it to the NSNO(0), confirming a Ni/Nd ratio close to 1 over the whole thin film. In addition, the elemental analysis demonstrates a possible degradation of the infinite-layer phase with a peculiar Nd/Ni off-stoichiometry at the surface of the thin film. The elongation of the *c*-axis parameter alongside the off-stoichiometry might suggest formation of spurious phase with a $Ni^{1+}$ varying towards higher valences near the surface of the thin film, because of a re-oxidation process. A zoom on the elemental EELS distribution at the film-substrate interface is shown in Figure 11d, and clearly demonstrates that the NSNO(0) is Nd terminated. The Sr EELS signal was not collected, but the Ti profile position is in it-self in agreement with a Ti-terminated STO, rendering a $NiO_2$-Nd-$TiO_2$-SrO interface reconstruction in perfect agreement with DFT calculation[38].

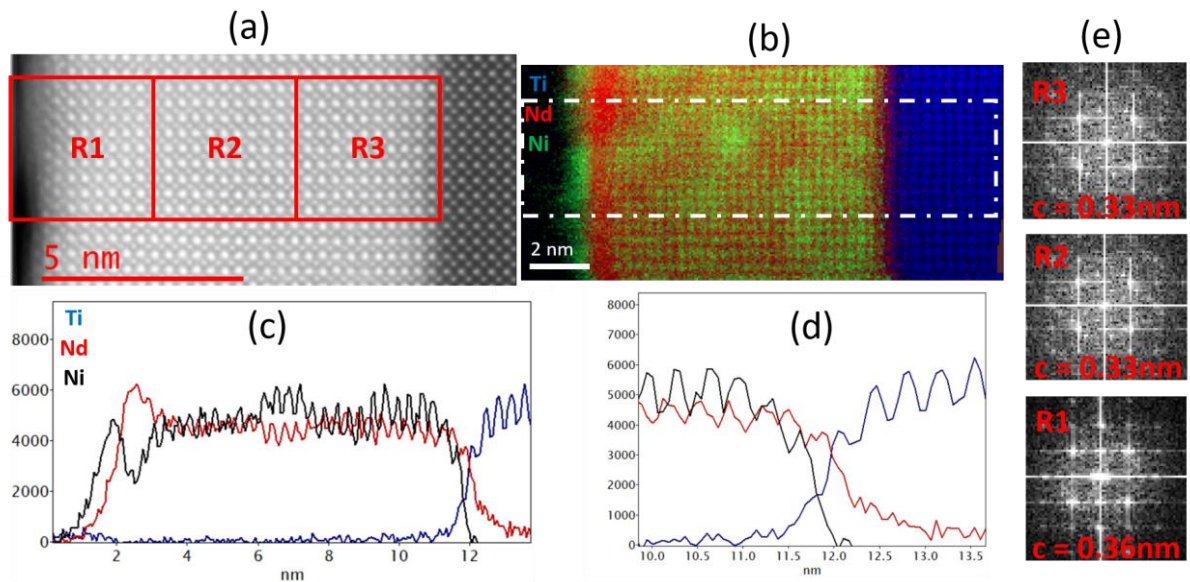

Figure 11 (a) HAADF-STEM image of a 10 nm NSNO(0) uncapped thin film after reduction (b) elemental STEM-EELS Nd, Ni, and Ti maps (c) elemental profiles integrated over 5 nm showing a Ni/Nd ratio close to 1 over the whole thin film, except for the surface and substrate/film interface for which one observes some segregation, (d) close-up on the elemental profiles at the NSNO(0)//STO interface, in good agreement with the $NiO_2$-Nd-$TiO_2$-SrO interface predicted by DFT calculations[38],(e) FFT of the three regions indicated by red boxes in (a), exhibiting *c*-axis parameters of 0.33 nm for R1 and R2, and 0.36 nm for R3.

## Anisotropic magnetoresistance measurements

Angle-dependent anisotropic magnetoresistance (AMR) measurements allowing the magnetic field (H) rotation for both in-plane (IP) and out-of-plane (OOP) geometries performed for reduced undoped $SrTiO_3$-capped and uncapped samples helped in delineating their intrinsic magnetic properties. Figure 12a sketches the H rotation with respect to the sample surface and the current I. The AMR amplitude is defined as: $[\rho(\varphi/\theta)-\rho(0)]/\rho(0)$. Figure 12b shows a clear fourfold $\varphi$-dependence for the IP-AMR data acquired at 2 K and 9 T for both capped and uncapped NSNO(0) samples. This fourfold symmetry is found to characterize both samples series down to 1 T. Below this value we did not observe any significant modulation of the resistance with the $\varphi$ angle. The IP-AMR magnitude is larger for the capped NSNO(0) samples, and we observe in both cases the maximum of the IP-AMR at $\pi/4$, along the Ni-Ni bonds, and the minimum at $\pi/2$ along the Ni-O-Ni bond. As shown in Figure 12c the IP-AMR magnitude decreases as the temperature is increased (fully vanishing above 20 K), and it increases as H is moved to higher values with a rather clear tendency to saturate, for both capped and uncapped samples.

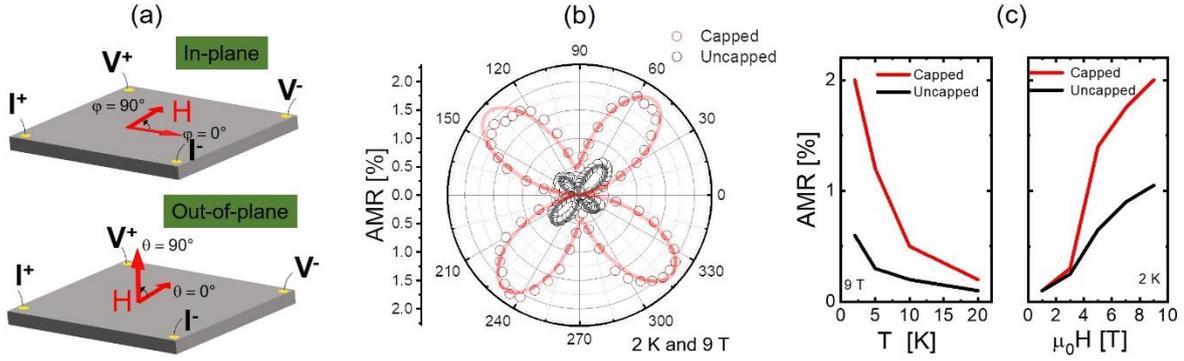

*Figure 12 (a) Transport configuration for in-plane (IP) and out-of-plane (OOP) angle-dependent AMR measurements. (b) Polar plots of the in-plane AMR as a function of the φ angle for measurements performed at 2 K under a magnetic field of 9 T for both reduced capped and uncapped NSNO(0) samples. Open dots represent the experimental data and solid lines the fitting curves. (c) Temperature- and magnetic field-dependences of AMR for both capped and uncapped reduced NSNO(0) samples.*

Figures 13a,b show the polar plot of the OOP-AMR data acquired at 2 K as a function of H for both capped and uncapped NSNO(0) samples. The OOP-AMR differ from the IP-AMR in the fact that they show an overall twofold symmetry. The magnitude of the signal is still larger for the capped samples, but these samples exhibit a non-monotonous H dependence, with a maximum around 4 T. On the contrary, the uncapped samples show an OOP-AMR intensity constantly increasing as a function of H, with a tendency to saturate at higher field values (upper panel of Fig. 13c), like the IP-AMR data.

A phenomenological formula was used to gain further insights on the AMR data, which accounted for both twofold and fourfold components[39] as indicated below:

$$\rho(\theta/\varphi) = A_2 \cdot \sin[2 \cdot (\theta/\varphi - p_2)] + A_4 \cdot \sin[4 \cdot (\theta/\varphi - p_4)] + C,$$

where $A_2$ and $A_4$ are the amplitudes of the twofold and fourfold terms, respectively, with $p_2$ and $p_4$ being their respective phases. C is a constant for a given set of H and temperature values. The $\theta$ and $\varphi$ angles are defined as sketched in Figure 12a. It was possible to satisfactorily fit the IP-AMR data with this formula, as shown in Figure 12b (solid lines). This fitting confirms that the fourfold $A_4$ component is indeed much more prominent for the capped samples. It was also possible to perfectly fit the OOP-AMR data acquired at 9 T and 2 K for both capped and uncapped NSNO(0) (solid lines in Figure 13d). The overall two-fold symmetry is largely modified by a feature appearing at $\theta = 0°$, 180° especially pronounced for the capped samples. From the fitting procedure, it results that the fourfold oscillation contribution $A_4$ is larger for the capped samples (as in the case of IP-AMR), and increases with H (lower panel of Figure 13c). The twofold contribution $A_2$ term, for its part, shows the same H dependence as the overall OOP-AMR for both capped and uncapped samples (upper panel of Figure 13c). The slight dropout at 9T evidences the increased relevance of the $A_4$ contribution for high fields.

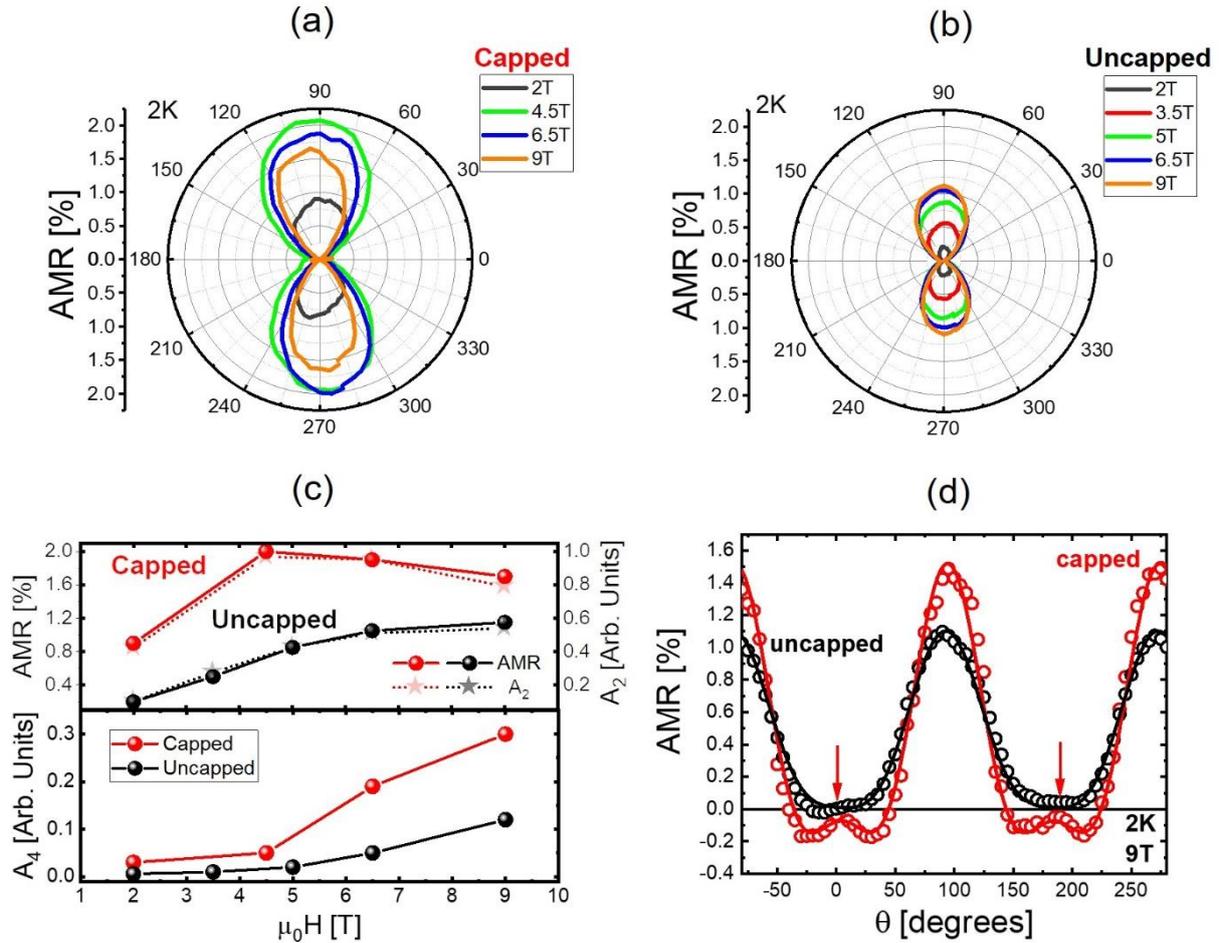

*Figure 13 Polar plots of OOP-AMR for (a) capped and (b) uncapped samples. (c) (Upper) Comparison of the twofold contribution $A_2$ parameter with the overall AMR as a function of the magnetic field for capped and uncapped NSNO(0). (Lower) Magnetic field dependence of the fourfold contribution $A_4$ parameter for capped and uncapped NSNO(0). (d) Appearing of a fourfold component (red arrows) on the AMR measured at 2K and 9T mostly prominent for the uncapped samples.*

AMR is a magnetotransport phenomenon where the measured resistivity variation depends not only on the strength of the applied magnetic field, but also on its relative orientation with respect to the current and crystal directions of the material[40]. AMR with two- or four-fold symmetry components and/or an admixture of them, have been already encountered in infinite-layer cuprates[41,65]. There, depending on the given system, the AMR is discussed in terms of different mechanisms encompassing the presence of antiferromagnetic long-range order[39,41,42], spin-orbit coupling[43–45], spin-charge segregation or stripes[46], spin-flop transition[47] or antiphase boundaries[48]. All these distinct explanations indicate a still non-reached consensus about the full understanding of this phenomenon in cuprates, while offering, nevertheless, a clear avenue in assessing their magnetic nature. *De facto*, the AMR phenomenology becomes more intriguing when one has to consider also the 4f magnetic moments of the $Nd^{3+}$ rare-earth sub-lattice, as in the case of NSNO(x) infinite-layer thin films. It has been already reported for (STO-capped) $Nd_{0.8}Sr_{0.2}NiO_2$ thin films that the presence of Nd-4f electrons largely modifies the magnitude of the superconducting upper critical fields, while resulting in AMR signals with complex angular dependences[49]. The observation for our undoped samples of a fourfold IP-AMR for H values ranging in the

1-9 T range, can be put in relation with the local $Nd^{3+}$ magnetic moments, which exhibit a planar easy configuration[49]. Indeed, in the case of electron-doped cuprates $Nd_{2-x}Ce_xCuO_4$ the IP-AMR signal showed a fourfold oscillation term[46,47,50,51], while for 4f-moments-less compounds such as $La_{2-x}Ce_xCuO_4$, the IP-AMR showed only a twofold symmetry[52]. It is clear that the presence of Kramers doublet $Nd^{3+}$ ions will modify the scattering mechanism of the charge carriers, especially in the very low temperature region and at high magnetic fields. We believe, on the specific case of infinite-layer nickelates, that this is uniquely possible because of an intra-atomic exchange coupling between the Nd-4f and Nd-5d orbitals as put forward theoretically[53]. This exchange coupling, by introducing a spin-disorder broadening of the $\Gamma$-centered electron pockets at the Fermi level, influences the scattering mechanism of the charge carriers, together with the Sr-doping level and the unique Nd5d-Ni3d hybridization[54]. The latter is the consequence of a non-zero spectral weight of the Nd-$5d_{z^2}$ bands at the Fermi level which creates electron pockets that self-localize at low temperature, and perhaps, explain the resistivity upturn for undoped and underdoped samples showed in Figures 10a,b. Those electron pockets are the result of a self-hole doping effect of the $NiO_2$ planes (up to 0.1 holes[55]), due to the transfer of electrons from the Ni $3d_{x^2-y^2}/3d_{z^2}$ states to the Nd 5d ones. First RIXS measurements in the low energy excitation region rendered spin excitations around 200 meV energy loss for STO-capped samples at different Sr-doping. The observed extra damping of the spin-waves upon increasing the Sr-content, indicates a decisive role played by those electron pockets[56], with (perhaps) consequences on the pairing mechanism itself. As a result, the presence of itinerant Nd-5d electrons marks an extra degree of freedom regarding the magnetic structure of the infinite-layer nickelates if compared to the one of hole-doped cuprates. The presented angle-dependent AMR measurements suggest that the magnetic properties of $NdNiO_2$ samples are also largely altered by the presence/absence of the STO capping-layer. We believe that the intra-atomic exchange coupling between Nd-4f and Nd-5d orbitals combined to the Sr-doping level and the aforementioned Nd-Ni hybridization that, on the other side, it has been clearly shown to be modulated by the presence/absence of a capping layer[17], may concomitantly control the AMR intensity, and therefore explain the magnetic properties of the infinite-layer nickelates as a function of the STO capping-layer. It remains to understand to which extent the Nd-Ni hybridization and/or the intra-atomic coupling between the Nd 5d and 4f states play a role in the intensity modulation of the IP(OOP)-AMR data, and in particular, of the mixed twofold/fourfold symmetry behaviour for the AMR data in the case of capped samples. Further experiments performed on samples synthesized with different capping layers, Sr-doping levels and rare-earth elements would perhaps contribute to allow a better understanding of those AMR measurements, which is out the scope of the present work.

**Conclusion**

We have reported about the synthesis of infinite-layer $Nd_{1-x}Sr_xNiO_2$ (x = 0, 0.05 and 0.2) thin films. Our optimized topotactic reduction processes for undoped and 5% at. Sr-doped $NdNiO_2$ STO-capped and uncapped thin films rendered very smooth and step-terraced surface morphologies which are a necessary and sufficient condition to perform surface-sensitive measurements, such as scanning tunnelling and/or ARPES, which are planned in the near future.

The obtained AMR data could be phenomenologically described by a model accounting for both twofold and fourfold symmetries. While overall XRD and resistivity data after topotactic reductions pointed at no significant differences between capped and uncapped samples, AMR curves for the undoped samples showed differences in amplitude and symmetry, therefore pointing to the fact that the crystal structure is by far not the only parameter of influence for this system. The capped sample showed larger AMR signals for both in-plane and out-of-plane configurations. They differ from the uncapped samples displaying, for high magnetic fields, a distinct four-fold symmetry contribution in addition to the mainly two-fold symmetry behaviour. We believe that our findings speak in favour of an influence of the STO capping-layer on the peculiar intra-atomic Nd 4f-Nd 5d exchange coupling and anisotropic Nd-Ni hybridization.

## Acknowledgements

This work was funded by the French National Research Agency (ANR) through the ANR-21-CE08-0021-01 'ANR FOXIES' and, within the Interdisciplinary Thematic Institute QMat, as part of the ITI 2021 2028 program of the University of Strasbourg, CNRS and Inserm, it was supported by IdEx Unistra (ANR 10 IDEX 0002), and by SFRI STRAT'US project (ANR 20 SFRI 0012) and ANR-11-LABX-0058_NIE and ANR-17-EURE-0024 under the framework of the French Investments for the Future Program. G.K. and D.P thank Dr. Jennifer Fowlie and Prof. H. Y. Hwang for useful discussions about the topotactic reduction process. The authors acknowledge the XRD platform of the IPCMS and Rene Baehr for technical support.

*daniele.preziosi@ipcms.unistra.fr